\documentstyle[prb,epsfig,aps,multicol]{revtex}
\begin{document}
\draft
 \title{
The correction-to-scaling exponent in dilute systems}
\author{
   {R.~Folk$^{1}$,
   Yu.~Holovatch$^{2}$,
   T.~Yavors'kii$^{3}$}}
\address{$^{1}$Institute for Theoretical Physics,
University of Linz, Linz, Austria \\
 $^{2}$Institute for Condensed Matter Physics,
Ukrainian Academy of Sciences,  Lviv, Ukraine \\
 $^{3}$Ivan Franko Lviv State University,
 Lviv, Ukraine}
\date{\today}
\maketitle
\begin{abstract}
The leading correction-to-scaling exponent $\omega$ for the
three-di\-men\-sional
dilute Ising model is calculated in the framework of the field 
theoretic renormalization group approach. Both in the minimal 
subtraction scheme as well as in the massive field theory (resummed  
four loop expansion) excellent agreement with recent Monte Carlo 
calculations [Ballesteros~H~G, et al Phys. Rev. B {\bf 58}, 2740  
(1998)] is achieved.  The expression of $\omega$ as series in a 
$\sqrt{\varepsilon}$-expansion up to ${\cal O}(\varepsilon^2)$ does 
not allow a reliable estimate for $d=3$.  \end{abstract}

\vspace{0.4cm}
\centerline{PACS: 64.60.Ak, 61.43.-j, 11.10.Gh }
\begin{multicols}{2}
From renormalization group (RG) theory one knows that in the
asymptotic region the values of the critical exponents are
universal and scaling laws between them hold. There the couplings of
the model Hamiltonian describing the critical system have reached 
their fixed point values. In the nonasymptotic region deviations from 
the fixed point values are present. They die out according to a 
universal power law governed by the correction-to-scaling exponent 
$\omega$.  E.g. for the zero field susceptibility the approach from 
above to the critical temperature $T_c$ is characterized by the 
so-called Wegner expansion \cite{Wegner72}
\begin{equation} \label{0}
\chi \simeq \Gamma_0 \tau^{-\gamma}\left(1+\Gamma_1\tau^{\omega/\nu}
+\Gamma_2\tau^{2\omega/\nu}+\ldots\right)\, ,
\end{equation}
where $\tau=(T-T_c)/T_c$ and the $\Gamma_i$ are the non-universal
amplitudes. $\gamma$ and $\nu$ are the asymptotic values
of the susceptibility and correlation length critical exponents.
The smaller the exponent $\omega$, the larger is the region where
corrections to the asymptotic power laws have to be taken into 
account.  Being even further away from the fixed point it is necessary 
to consider the complete non linear crossover functions.

The implication of quenched dilution on the critical behavior is a
long-standing problem attracting theoretical, experimental and 
numerical efforts. In the $3d$-Ising model quenched disorder changes 
the asymptotic critical exponents compared to the pure  ones 
\cite{Harris74,Chayes86}.  In principle this statement should hold for 
arbitrary weak dilution. But in order to observe this change one 
should approach the critical point close enough.  The width of this 
region turns out to be dilution dependent.

In particular Monte Carlo (MC) calculations of the critical exponents 
in the dilute $3d$-Ising model are more difficult to perform than for 
the pure model since they need much larger sizes of lattices 
\cite{Landau80}.  Even then the exponents were found to be 
non-universal and varying continously with dilution, i.e. they were 
effective ones \cite{Wang89}. It became clear that a 
correction-to-scaling analysis is unavoidable and indeed universal 
exponents were found \cite{Parisi98}. Without it one still obtains 
concentration dependent effective exponents \cite{Wiseman98}.

The value of the correction-to-scaling exponent $\omega$ found in MC
calculations from an analysis invoking the first correction
term in (\ref{0}) turned out to be \cite{Parisi98}
\begin{equation} \label{1}
\omega=0.37\pm 0.06  \, .
\end{equation}
Thus it is almost half as large as its corresponding value in
the pure model (see Table \ref{tab1}) and this smallness of $\omega$
in the dilute case explains its importance for an analysis of the
asymptotic critical behavior. It is therefore highly desirable to
have an independent quantitative theoretical prediction for the
value of the correction-to-scaling exponent in the dilute system.

In theoretical calculations the value of $\omega$ found by scaling 
field RG \cite{Newman82} is $\omega=0.42$.  So far field theoretical 
RG studies mainly concentrated on the asymptotic values of the leading 
exponents. Correction-to-scaling exponents
\narrowtext
 \begin{table} [b]
\begin{tabular}{l|l|l}
Method              & dilute                  &    pure \\
\hline
scaling field      & $0.42$\cite{Newman82}     & $0.87$\cite{Newman82}   
\\ $\varepsilon$ expansion & see text & $0.814\pm 0.018$ 
\cite{Guida98}   \\ massive RG, $d=3$     & $0.372\pm 0.005$   & 
$0.799\pm 0.011$ \cite{Guida98}  \\ min. sub. RG, $d=3$ &   $0.390\pm 
0.04$      &  $0.791\pm 0.036$ \\ &                        & 
$0.8\pm0.1$\cite{Schloms};  \\ MC        &  $0.37\pm 
          0.06$\cite{Parisi98}   &  $0.8-0.85$\cite{Baillie92}; \\ &                                 
          & $0.87\pm 0.09$ \cite{Ballesteros98} 
\end{tabular}
\caption{\label{tab1} Values of correction-to-scaling exponent 
$\omega$ as obtained from different methods in dilute and pure 
$3d$-Ising models.} 
\end{table} 
have been calculated within massive RG in two loop approximation in 
Ref. \cite{Jug83} ($\omega=0.450$)  and within the minimal subtraction 
scheme in three loop approximation in Ref. \cite{Janssen95} 
($\omega=0.366$). Here, we improve this value in the massive RG scheme 
up to four loop  order with the result \begin{equation} \label{1a} 
\omega=0.372\pm 0.005 \end{equation} in excellent agreement with 
(\ref{1}).  In the minimal subtraction scheme we obtain 
$\omega=0.390\pm 0.04$ remaining with in the bandwidth of MC accuracy.

The critical behavior of the quenched weakly dilute
Ising model in the Euclidian space of $d=4-\varepsilon$ dimensions is
governed by a Hamiltonian with two couplings \cite{Grinstein76}:
\begin{eqnarray}
\label{2}
{\cal H}(\phi)&=&\int {\rm d}^dR \Big\{ {1\over 2} \sum_{\alpha=1}^{n}
\left[|\nabla {\phi}_\alpha|^2+ m_0^2 {\phi}_\alpha^2\right]
\nonumber \\ &-&{v_{0}\over
4!} \left(\sum_{\alpha=1}^{n}{\phi}_\alpha^2 \right)^2 +
 {u_{0}\over 4!} \sum_{\alpha=1}^{n}{\phi}_\alpha^4 \Big\},
\end{eqnarray}
in replica limit $n \rightarrow 0$.  Here
$\phi_\alpha$ are the components of order parameter;
$u_{0} > 0, v_{0} > 0$ are bare couplings; $m_0$ is bare mass.

We describe the long-distance properties of the model (\ref{2})
in the vicinity of the phase transition point using a 
field-theoretical RG approach. The results presented in this paper are 
obtained on the basis of two different RG schemes: the normalization 
conditions of massive renormalized theory at fixed \cite{Parisi80} 
$d=3$  and the minimal subtraction scheme \cite{tHooft72}. The last 
approach allows both fixed $d=3$ calculations \cite{Schloms} as well 
as an $\varepsilon$-expansion.

In the RG method the change of  the couplings $u$ and $v$ under 
renormalization is described by two $\beta$-functions
\begin{eqnarray}  \label{3}
\beta_u(u,v) &=& \mu \left(\frac{\partial u}{\partial \mu}\right)_0,
\nonumber \\
\beta_v(u,v) &=& \mu \left(\frac{\partial v}{\partial \mu}\right)_0 \, ,
\end{eqnarray}
where $\mu$ corresponds to the mass in the massive field theory 
approach and to the scale parameter in the minimal subtraction scheme. 
The subscript in (\ref{3}) indicates that the derivatives are taken at
constant unrenormalized parameters.
The $\beta$-functions differ for different RG schemes and in 
consequence the fixed point coordinates $u^*$, $v^*$, defined by the 
simultanious zeros of both $\beta$-functions, are scheme dependent. 
The asymptotic critical exponents as well as the correction-to-scaling 
exponent do not depend on the RG scheme and take universal values.

The correction-to-scaling exponent $\omega$ is defined by the 
smallest eigenvalue of the matrix of derivatives of the 
$\beta$-functions
\begin{equation}  \label{4}
\pmatrix{\frac{\partial \beta_u}{\partial u}&\frac{\partial 
\beta_u}{\partial v} \cr     \frac{\partial \beta_v}{\partial 
u}&\frac{\partial \beta_v}{\partial v}\cr} 
\end{equation} 
taken at the stable fixed point. For the stable fixed point both 
eigenvalues of this matrix have a positive real part.

Our results for the correction-to-scaling exponent are based on the 
known high order expansions for the functions $\beta_u$ and $\beta_v$. 
In the massive scheme they are known in four loop approximation 
\cite{Mayer89}. In the minimal subtraction scheme one can obtain these 
functions in five loop approximation in the replica limit from those 
of a cubic model \cite{Kleinert95}. In the limiting case of the pure 
model only the coupling $u$ is present. The corresponding
$\beta$-function results from putting $v=0$ in $\beta_u(u,v)$  and the
correction-to-scaling exponent is simply the derivative $\partial 
\beta_u(u,0)/ \partial u $ taken at the stable fixed point $u^*$. Note 
that for the pure model the $\beta$-functions in the massive scheme 
are known in six loop approximation \cite{Baker78} and the five loop 
results for the RG functions in the minimal subtraction scheme 
\cite{Kleinert91} agree with those recovered from Ref. 
\cite{Kleinert95}.

It is known that the series obtained in the perturbational RG approach 
are at best asymptotic (for the dilute model see however Ref. 
\cite{Bray87}). An appropriate resummation procedure has to be applied 
to the $\beta$ functions in order to obtain reliable information. The 
choice of the resummation procedure depends on the information about 
the high order behavior of the expansion series. This information is 
not available for the case of the $\beta$-functions (\ref{3}). In this 
situation the most appropriate way to proceed, is to use the 
Pad\'e-Borel resummation \cite{Baker78} generalized for the two 
variable case \cite{note1}.

The steps which we follow in the calculation of the 
correction-to-scaling exponent $\omega$ are the following: First the 
$\beta$-functions (\ref{3}) are resummed and the system of equations 
for the fixed points, $\beta_u(u^*,v^*)=0$, $\beta_v(u^*,v^*)=0$, is 
solved. Then the matrix of derivatives (\ref{4}) is calculated for the 
resummed $\beta$-functions.  The stability of the fixed points is 
checked. The fixed point with both $u^*\neq 0$ and $v^*\neq 0$ is the 
stable one at $d=3$ and the smallest eigenvalue gives the desired 
correction-to-scaling exponent.  Note that the eigenvalues might be 
complex, in this case both have the same positive real part defining 
$\omega$.

\begin{figure}[h,b]
       \epsfig{file=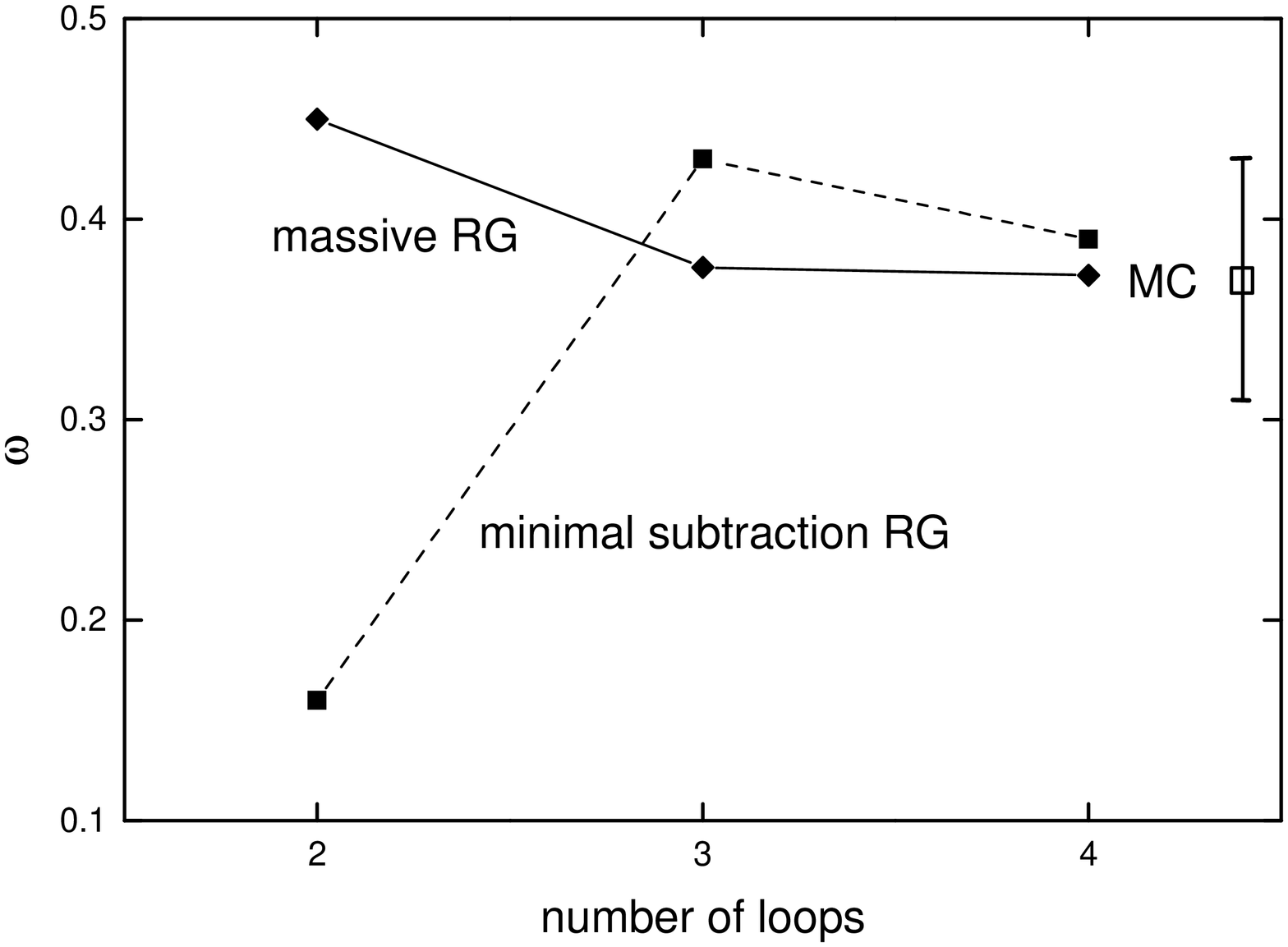,height=8cm,width=6cm,angle=-90}
    \caption{\label{fig1} Correction-to-scaling exponent
$\omega$ of the dilute 3d Ising model in increasing number of loops. 
Open square with error bar shows the region of accuracy of the MC data 
\protect\cite{Parisi98}; full squares: our values in the minimal 
subtraction RG scheme; full diamonds: our values in the massive RG 
  scheme.} \end{figure}

In Fig. \ref{fig1} we present our results for the exponent $\omega$
obtained in successive orders of perturbation theory in number of 
loops.  To perform the resummation the Borel transforms of the 
truncated $l$th order perturbation theory expansion for the 
$\beta$-functions were presented in the form of $[(l-1)/1]$ rational 
approximants of two variables \cite{note1}.  This form of rational 
approximants appeared to give the most reliable results.  The four 
loop results for the exponent $\omega$ obtained in both RG schems are 
given in the second column of Tab. \ref{tab1}. For the uncertainty of 
the four loop results we simply take the difference between the four 
loop and the three loop result. This is suggested by the behavior of 
$\omega$ in succesive numbers of loops shown in Fig. \ref{fig1}. 
Although both RG schemes lead to comparable values for $\omega$, the 
convergence of the values in the massive scheme is much faster 
(however the accidentally very small error for $\omega$ derived from 
this procedure does not hold for other quantities).  Note that the 
result for $\omega$ combined with the corresponding four loop results 
for the asymptotic critical exponents \cite{Mayer89,Folk98} confirms 
the conjectured inequality, $-\nu\omega <\alpha < 0$, for the random 
models critical exponents involving the specific heat exponent 
$\alpha$ \cite{Aharony98}.

\narrowtext

As it was noted above five loop results for the minimal subtraction 
scheme are available \cite{Kleinert95}. In particular applying the 
resumation scheme \cite{note2} to the pure Ising model case, $v=0$, we 
get the following values for $\omega$ in increasing number of loops 
starting from two loop: $\omega =0.566;\, 0.852; \, 0.756; \, 0.791$. 
This leads to an improvement in accuracy of the previously calculated 
$d=3$ five loop value \cite{Schloms} since $\omega= 0.791\pm 0.036$, 
the value and its uncertainty now is comparable to the six loop 
calculation within the massive scheme \cite{Guida98} (see the third 
column of Tab. \ref{tab1}).

The degeneracy of the dilute Ising model $\beta$-functions on the one 
loop level leads to the $\sqrt{\varepsilon}$-expansion 
\cite{Grinstein76,Khmel75}. For the critical exponents this expansion 
is known  up to ${\cal O}(\varepsilon^2)$ \cite{Shalaev97}. Starting 
from the five loop results of Ref. \cite{Kleinert95} in the replica 
limit we get the following expansions \cite{note4} for the eigenvalues 
$\omega_1$ and $\omega_2$ of the stability matrix (\ref{4}) in the 
fixed point $u^*\neq 0$, $v^*\neq 0$:
\begin{eqnarray} \label{5}
\omega_1&=&2\, \varepsilon+3.704011194 \,
\varepsilon^{3/2}+11.30873837 \,\varepsilon^2 \, ,
\nonumber \\  \nonumber
\omega_2&=&0.6729265850\, \varepsilon^{1/2}-1.925509085 \, 
\varepsilon  \\ &-&0.5725251806 \, \varepsilon^{3/2} -13.93125952 \, 
\varepsilon^2 \, .  
\end{eqnarray}
From naively adding the successive perturbational contributions one 
observes that already in three loop approximation ($\sim \varepsilon$) 
$\omega_2$ becomes negative and therefore {\bf no stable fixed point} 
exists in strict $\sqrt{\varepsilon}$-expansion. Even the resummation 
procedures we applied above, do not change this picture \cite{Folk98}. 
This can be considered as indirect evidence that the 
$\sqrt{\varepsilon}$-expansion is not Borel summable, as may be 
expected from Ref. \cite{Bray87}. A physical reason might be the 
existence of the Griffith singularities caused by the zeros of the 
partition function of the pure system \cite{Griffith69}. The fixed $d$ 
approach, both within the massive \cite{Parisi80} and minimal 
subtraction \cite{tHooft72,Schloms} schemes, seems to be the only 
reliable way to study critical behaviour of the model by means of RG 
technique.

\acknowledgments{We acknowledge valuable correspondance with
Alan J. McKane  and Victor Mart\'in-Mayor.}

\end{multicols}


\begin{thebibliography}{99}
\bibitem{note1} Here we use the Chisholm-Borel resummation technique 
for a series of two variables (see \cite{Jug83}).
\bibitem{note2} In this case the Pad\'e-Borel resummation based on
$[(l-1)/1]$ Pad\'e approximant for the Borel transform is recovered.
\end{thebibliography}
\end{document}